\documentclass[aps,pra,10pt,amsmath,twocolumn,notitlepage,tightenlines,superscriptaddress,floatfix,longbibliography]{revtex4-1}

\usepackage{bm}
\usepackage{mathtools}
\usepackage{graphicx}
\usepackage[caption=false, subrefformat=parens, labelformat=parens]{subfig}
\usepackage{xcolor}
\definecolor{dark-red}{rgb}{0.4,0.15,0.15}
\definecolor{dark-blue}{rgb}{0.15,0.15,0.4}
\definecolor{medium-blue}{rgb}{0,0,0.5}
\usepackage{hyperref}
\usepackage[all]{hypcap}
\usepackage{cleveref}
\hypersetup{
    colorlinks, linkcolor={dark-red},
    citecolor={dark-blue}, urlcolor={medium-blue}
} 

\newcommand{\nsp}{\hspace{-0.4pt}}
\newcommand{\ssp}{\hspace{0.4pt}}
\newcommand{\norm}[1]{\lvert #1 \rvert}
\newcommand{\normb}[1]{\big\lvert #1 \big\rvert}
\newcommand{\proj}[1]{\ket{#1}\nsp \bra{#1}}
\newcommand{\ket}[1]{\lvert\, #1\, \rangle}
\newcommand{\ketb}[1]{\big\lvert\, #1\, \big\rangle}
\newcommand{\bra}[1]{\langle\, #1\, \rvert}
\newcommand{\brab}[1]{\big\langle\, #1\, \big\rvert}
\newcommand{\braket}[2]{\langle\, #1\,\vert\, #2 \,\rangle}
\newcommand{\half}{\frac{1}{2}}
\newcommand{\bms}{{\bm s}}
\newcommand{\bmu}{{\bm u}}
\newcommand{\BB}{{b}}
\newcommand{\WW}{{w}}

\newcommand{\orderO}{{O}}
\newcommand{\xx}{{\varphi}}

\newcommand{\ea}{{\it et al.}}

\begin{document}

\title{Near-optimal quantum circuit for Grover's unstructured search using a transverse field}
\date{\today}
\author{Zhang Jiang}
\email{zhang.jiang@nasa.gov}
\affiliation{Quantum Artificial Intelligence Laboratory (QuAIL), NASA Ames Research Center, Moffett Field, California 94035, USA}
\affiliation{Stinger Ghaffarian Technologies Inc., 7701 Greenbelt Rd., Suite 400, Greenbelt, MD 20770}

\author{Eleanor G. Rieffel}
\affiliation{Quantum Artificial Intelligence Laboratory (QuAIL), NASA Ames Research Center, Moffett Field, California 94035, USA}

\author{Zhihui Wang}
\affiliation{Quantum Artificial Intelligence Laboratory (QuAIL), NASA Ames Research Center, Moffett Field, California 94035, USA}
\affiliation{Universities Space Research Association, 615 National Ave, Mountain View, CA 94043}

\begin{abstract}
Inspired by a class of algorithms proposed by Farhi 
\ea~(arXiv:1411.4028), namely the quantum
approximate optimization algorithm (QAOA), we present a circuit-based quantum
algorithm to search for a needle in a haystack, obtaining the same quadratic
speedup achieved by Grover's original
algorithm. In our algorithm, the problem Hamiltonian (oracle) and
a transverse field 
are applied alternately to the system in a periodic manner. We introduce a technique, based on spin-coherent states, to analyze the composite unitary in a single period. This composite unitary drives a closed transition
between two states that have high degrees of overlap with the initial state 
and the target state, respectively. 
The transition rate in our algorithm is of order
$\Theta(1/\sqrt N)$, and the overlaps are of order $\Theta(1)$, yielding a
nearly optimal query complexity of $T\simeq \sqrt N\ssp (\pi/2\sqrt 2\,)$. 
Our algorithm is a QAOA circuit that demonstrates a quantum advantage with a large number of iterations that is not derived from Trotterization of an adiabatic quantum optimization (AQO) algorithm. It also suggests that the analysis required to understand QAOA circuits involves a very different process from estimating the energy gap of a Hamiltonian in AQO.

\end{abstract}

\maketitle

\section{Introduction}
\label{sec:intro}

Recently, Farhi \ea~\cite{farhi_quantum_2014, farhi_quantum_2014-1} proposed a
new class of quantum heuristic algorithms, the quantum approximate 
optimization algorithm (QAOA). 
We present an algorithm for Grover's unstructured search problem 
\cite{grover_fast_1996} inspired by QAOA.
This algorithm shows a quantum advantage for a 
number of iterations $p$ in the intermediate range between $p=1$ and $p\to\infty$. 
We also introduce a tool, a representation based on spin-coherent states,
for the design and analysis of the QAOA-type circuits. Using this tool, 
we prove a $\Theta(\sqrt N)$ query complexity for our algorithm. 
The algorithm has the advantage of requiring fewer two-qubit gates
than Grover's original algorithm because we use the transverse field in 
place of Grover's original diffusion operator. 
With an increasing number of iterations $p$, an exhaustive search of the QAOA
parameters often becomes inefficient due to the curse of dimensionality. Our
method avoids this difficulty by restricting the parameters to be periodic.
The approach suggests a potential route for parameter optimization 
for QAOA-based quantum heuristic algorithms more generally. 

In our algorithm, mixing and problem (oracle) Hamiltonians are applied 
to the system in a sequence that is periodic in time. The long-time 
dynamics of a periodically driven quantum system can be
profoundly different from a time-homogeneous
one~\cite{goldman_periodically_2014}. To analyze the outcome after
$\Theta(\sqrt N)$ periods, we solve the relevant eigenvalues and eigenvectors
of the composite (effective) unitary in a single period to exponential
precision $\orderO(1/\sqrt N)$. This analysis gives further evidence that,
while the initial motivation for Farhi \ea's design of 
QAOA circuits may have come from Trotterization 
of adiabatic quantum optimization (AQO), the analysis required to 
understand QAOA circuits involves a very different process from 
estimating an exponentially small energy gap of a Hamiltonian. 

Instead, the intuition for this algorithm comes from a 
phase-space representation based on spin-coherent states in
which both the unitaries generated by the mixing and the oracle Hamiltonians
take simple forms. We find that the composite unitary generates a closed
transition between two states that have high degrees of overlap with the
initial state 
and the target state, respectively. 
The transition rate in our algorithm is of order $\Theta(1/\sqrt N)$, 
and the overlaps are of order $\Theta(1)$, yielding a nearly optimal query
complexity of $T\simeq \sqrt N\ssp (\pi/2\sqrt 2\,)$. 


We begin, in Sec.~\ref{sec:QAOA}, by briefly reviewing QAOA circuits, providing
context and inspiration for our construction.
In Sec.~\ref{sec:grover}, we briefly review prior approaches to Grover's 
problem.
In Sec.~\ref{sec:main}, we introduce our algorithm. 
Section~\ref{sec:scs} gives an intuitive picture, 
using a representation based on spin-coherent states, 
for why the algorithm works.
The most straightforward application of this picture results in a query
complexity that is close to optimal, up to a polylog factor. We then
refine the algorithm, removing the polylog factor, to obtain a query
complexity within a small constant of the optimal value. This improvement
makes use of the  phase-space representation we describe 
in Sec.~\ref{sec:phase_space}.  
Section~\ref{sec:eigen_V} shows how we use 
this phase space representation to derive
analytical results, including the success probability and the query complexity
of our algorithm. In Sec.~\ref{sec:check}, we briefly comment on how to check
whether the correct solution has been found. We conclude in 
Sec.~\ref{sec:conclusion} with thoughts on future directions.

\section{Review of QAOA circuits}
\label{sec:QAOA}

QAOA circuits iteratively alternate between a classical 
Hamiltonian (usually the problem Hamiltonian derived from a cost function) 
and a mixing term (often the transverse field)~\cite{farhi_quantum_2014, farhi_quantum_2014-1}. Farhi \ea\ 
proposed these circuits to tackle
approximate optimization of challenging combinatorial problems, with
the approximation ratio improving (or at least not decreasing) 
as the number of iterations $p$ increases. We will refer to circuits with the above structure as QAOA circuits
whether or not they are used for approximate optimization or for some other
purpose. Since Farhi \ea's original work, 
QAOA circuits have also been applied for exact optimization
\cite{wecker2016training} and sampling \cite{farhi_quantum_2016}.
Further, Farhi and Harrow~\cite{farhi_quantum_2016} argued, under 
reasonable complexity theoretic assumptions, that 
it is not possible for any classical algorithm to produce samples according
to the output distribution of QAOA circuits with even a single iteration ($p = 1$).
Their results suggest that QAOA circuits applied to sampling are 
among the most promising candidates for early
demonstrations of ``quantum supremacy''~\cite{preskill_quantum_2012,
boixo_characterizing_2016}.  
It remains an open question whether QAOA circuits provide a quantum
advantage for approximate optimization.

Trotterization of adiabatic quantum optimization (AQO) implies that
QAOA can always achieve the optimum in the limit of infinite iterations ($p\to\infty$). At the other end of the
spectrum, Farhi \ea~\cite{farhi_quantum_2014-1} proved that a QAOA circuit
with $p = 1$ beat the best classical approximation ratio for MaxE3Lin2 (each constraint is a linear equation mod 2 on 3 variables) at
the time; this quantum circuit then inspired a new classical approach that currently hold the
record~\cite{barak_beating_2015_published}). 
The parameters for these circuits are the times $\beta_i$ and $\gamma_i$,
$1 \leq i \leq p$,
for which the mixing and classical Hamiltonian, respectively, are applied.
Farhi \ea\ show that, for a fixed $p$, the optimal parameters can be computed
in polynomial time in the number of qubits $n$. If we discretize so that each parameter can take on $m$ values, an exhaustive search for the optimum takes exponential steps in $p$ as $m^{2p}$.

For this reason, prior to this work, 
there were no results for QAOA circuits with an intermediate number of 
iterations $1 \ll p < \infty$. Here, we give such an algorithm. 
Our approach suggests that considering QAOA
circuits with periodic parameters may be a profitable way for parameter
setting for QAOA circuits with $1 \ll p < \infty$.


\section{Review of prior quantum algorithms for Grover's problem}
\label{sec:grover}


Grover's algorithm~\cite{grover_fast_1996}
has attracted much attention, because it has been proven
that it outperforms any classical algorithm. 
It searches for a needle in a haystack, achieving
a query complexity of $\Theta(\sqrt N)$, 
where $N = 2^n$ is the size of the search space. 
Grover's algorithm is optimal among quantum algorithms for such a
task~\cite{bennett_strengths_1997, farhi_analog_1998, zalka_grovers_1999}.
It offers a modest quadratic speedup over any classical
counterpart, although even quadratic speedup is considerable when $N$ is large.

Grover's algorithm selectively alters the phase of the target state given by
the oracle, at each iteration. While this operation on its own would not change
the probability of reading out the target state, it sets the stage for the 
next operation which takes advantage of the phase difference 
to increase the probability of the system being in that state. This effect
would be impossible were quantum amplitudes not able to store 
phase information as well as the probability. 
This step is carried out by Grover's diffusion operator, which
applies a phase of $\pi$ to the even superposition state 
and does nothing to any state orthogonal to it. 
It requires $\Theta(n)$ two-qubit gates to
implement Grover's diffusion operator~\cite{diao_quantum_2002}. 

Grover's unstructured search problem can also be solved by adiabatic quantum
computation, where a mixing Hamiltonian (typically a transverse field) is
gradually replaced by the problem Hamiltonian that encodes the answer in
its ground state. The minimum gap of the total Hamiltonian is crucial to the
time complexity of the algorithm and was first given by Farhi 
\ea~\cite{farhi_quantum_2000}. 
Recently, the exponential scaling of this minimum gap was rederived using an
instanton approach, without solving the eigenvalue equation (see Supplemental
Material in~\cite{isakov_understanding_2016}).  
By adjusting the evolution rate of the
Hamiltonian, Roland and Cerf~\cite{roland_quantum_2002} recover the quadratic
advantage of Grover's original algorithm over classical search.
Roland and Cerf do not use the standard mixing operator, the transverse 
field, but rather a Hamiltonian related to Grover's diffusion operator. 

A natural question
is whether it is possible to implement unstructured quantum search in
the circuit model using the transverse field instead of Grover's diffusion
operator. Here, we give an affirmative answer to this question.

\section{Our Algorithm}
\label{sec:main}

Here, we give a high-level view of the algorithm.
Sec.~\ref{sec:scs} describes the intuition behind our algorithm, based 
on a picture using spin-coherent states. 

{\it Grover's problem.} Suppose we are given a problem Hamiltonian (oracle)
\begin{align}
  C_{\bmu} = - \proj{\bmu}\,,
\end{align}
that encodes an unknown bit string $\bmu$ of length $n$ ($n$ is even, for
simplicity).  The aim is to find $\bmu$ using as few calls to this
oracle as possible. 

Our algorithm uses the transverse field operator $B$ as the driver 
(mixing term),
\begin{align}
B =\sum_{j=1}^n X_j\,,
\end{align}
where $X_j$ is the Pauli $X$ operator of the $j$th qubit. An advantage of
using $B$ over Grover's diffusion operator is that $B$ acts only on
individual spins, so it is easier and more efficient to implement. 
The input state of
our algorithm is the usual one, the tensor product $\ket{+}^{\otimes n}$, the 
joint $+1$ eigenstate of all the $X_j$ operators, and the even superposition
of all bit strings, 
\begin{align}\label{eq:initial}
 \ket{\psi_\mathrm{in}} =  \ket{+}^{\otimes n} = \frac{1}{\sqrt N} \sum_{\bms\in \{0,1\}^n} \ket{\bms}  \,.
\end{align}

We can simplify the analysis, following Farhi \ea~\cite{farhi_quantum_2000},
by working in a basis in which the target state is $\ket{\bm 0} = \ket{0\cdots 00}$.
Since the driver $B$ and the initial state $\ket{\psi_\mathrm{in}}$ remain the
same when any subset of the $n$ qubits is flipped, the problem can be
converted to finding the bit string $\bm 0$ using the oracle $C_{\bm 0}$ with
the same driver $B$. Doing so drastically simplifies our analysis: 
the state $\ket{\bm 0}$ and the initial state $\ket{+}^{\otimes n}$
are in the $(n + 1)$-dimensional symmetric subspace (under permutations of 
qubits), and the evolution under both $B$ and $C_{\bm 0}$ preserves this 
subspace, so we need to consider only that $(n + 1)$-dimensional subspace 
instead of the whole Hilbert space of dimension $2^n$. 
To simplify notation, we will omit the
subscript in $C_{\bm 0}$ hereafter, i.e., $C\equiv C_{\bm 0}$.

\begin{figure}
\label{fig:grover_circuit}
\includegraphics[width=0.48\textwidth]{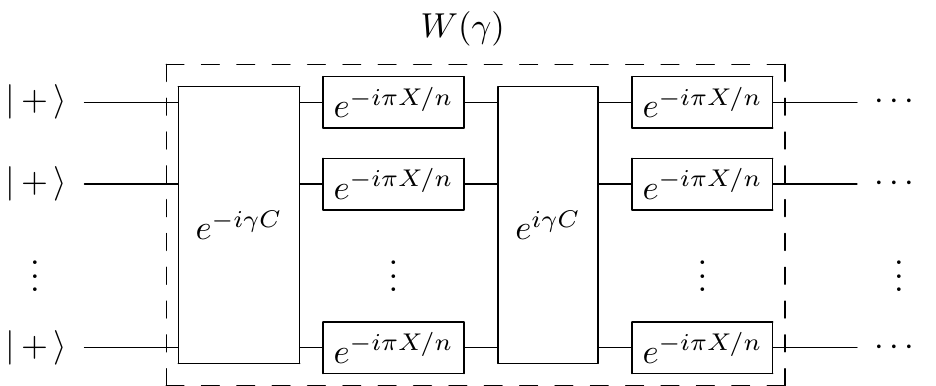}
\caption{To map the input state to a state having large overlap with the
target, the unitary $W(\gamma)$ is repeated for $\orderO(N^{1/2})$ times.}
\end{figure}

The building block of our algorithm is a simple product of 
unitaries generated by $B$ and $C$,
\begin{align}\label{eq:W}
 W(\gamma) = e^{-i\pi B/n}e^{i\gamma C}e^{-i\pi B/n}e^{-i\gamma C}\,,
\end{align}
where $\gamma \in (0, \pi]$ is a free parameter. The intuition for why we choose the angle of the rotation $e^{-i\pi B/n}$ can be found in Sec.~\ref{sec:scs}. 
The algorithm repeatedly applies the unitary $W(\gamma)$ for 
$\Theta(\sqrt N)$ times (see Fig.~\ref{fig:grover_circuit}). 
The relevant eigenvalues of the unitary $W(\gamma)$ determine the query 
complexity of our algorithm, while the corresponding eigenvectors determine 
the probability of success.  
We will show that the relevant eigenvalues are the ones closest to $1$, but
not equal to $1$.

The unitary $W(\gamma)$ has a time-reversal-like symmetry
\begin{align}\label{eq:time_reversal}
 \Lambda\ssp W(\gamma) \Lambda^\dagger = W^\dagger(\gamma)\,,
\end{align}
where $\Lambda = e^{-i\pi B/n} Z_1 Z_2\cdots Z_n$ with $Z_j$ being the
Pauli-$Z$ operator of the $j$th qubit. Equation~(\ref{eq:time_reversal})
holds generally for Hamiltonians based on classical cost functions, 
Hamiltonians diagonal in the computational basis.
This symmetry implies that if
$\alpha$ is an eigenvalue of $W(\gamma)$, then its complex conjugate 
$\alpha^*$ is also an eigenvalue
of $W(\gamma)$; the corresponding eigenstates are denoted by $\ket{\WW_\alpha}$
and $\ket{\WW_{\alpha^*}}$, respectively. 
When restricted to the two-dimensional subspace $\mathcal S_\alpha$ spanned
by $\{\ket{\WW_\alpha}, \ket{\WW_{\alpha^*}}\}$ and written in the
basis $\{\ket{\WW_+}, \ket{\WW_-}\}$, where
\begin{align}
 \ket{\WW_\pm} = \frac{1}{\sqrt 2}\Big(\ket{\WW_\alpha} \pm \ket{\WW_{\alpha^*}}\Big)\,,
\end{align}
$W(\gamma)$ has the matrix representation
\begin{align}
 W\big|_{\mathcal S_\alpha}(\gamma) = 
  \exp\left[\mathord{-}i\left(\nsp
  \begin{matrix}
   0 & \arg(\alpha)\\
   \arg(\alpha) & 0
  \end{matrix}\right)\right]\,.
\end{align}
The unitary $W(\gamma)$ thus
drives a closed transition between $\ket{\WW_\pm}$ with the transition rate
$\arg(\alpha)$. To drive a full transition, one needs to repeat $W(\gamma)$ for
roughly $\pi/[2\arg(\alpha)]$ times.

Let $\ket{\BB_\pm} = \frac{1}{\sqrt 2}\big(\ssp\ket{+}^{\otimes n} \pm \ket{-}^{\otimes n}\big)$. 
We show in Sec.~\ref{sec:eigen_V} that for eigenvalues $\alpha$ and $\alpha^*$ exponentially close to $1$ but not equal to
$1$, $\ket{\WW_\alpha}$ and $\ket{\WW_{\alpha^*}}$ have large overlaps with
$\frac{1}{\sqrt 2}\big(\ssp\ket{\bm 0}\pm i \ket{\BB_+}\big)$, respectively. In
other words, $\ket{\WW_+}$ and $\ket{\WW_-}$ have large overlaps with $\ket{\bm
0}$ and $i \ket{\BB_+}$, respectively, so the algorithm drives $\ket{\BB_+}$
close to the target state $\ket{\bm 0}$. 
The value of $\arg(\alpha)$ has to be exponentially small in $n$; otherwise,
our algorithm would have beaten the optimal query complexity of Grover's
algorithm.  Hereafter, $\alpha$ will refer to this specific eigenvalue. 
The initial state~(\ref{eq:initial}) can be written as
\begin{align}
 \ket{\psi_\mathrm{in}} = \ket{+}^{\otimes n} = \frac{1}{\sqrt 2}\Big(\ket{\BB_+} + \ket{\BB_{-}}\Big)\,;
\end{align}
note that $\ket{\BB_{-}}$ is a dark state, i.e.,
$W(\gamma) \ket{\BB_-}=\ket{\BB_{-}}$. For
$\norm{\braket{\bm 0}{\WW_+}}\simeq \norm{\braket{\BB_+}{\WW_-}}\simeq 1$, 
the output state is approximately
\begin{align}
 \ket{\psi_\mathrm{out}} \simeq \frac{1}{\sqrt 2}\Big(\ket{\bm 0}+\ket{\BB_-}\Big)\,,
\end{align}
and the probability of finding the target state $\ket{\bm 0}$ is approximately $1/2$. 

\begin{figure}
\subfloat[]{\hspace{-1.5mm}
\includegraphics[width=.23\textwidth, height= 3.3cm]{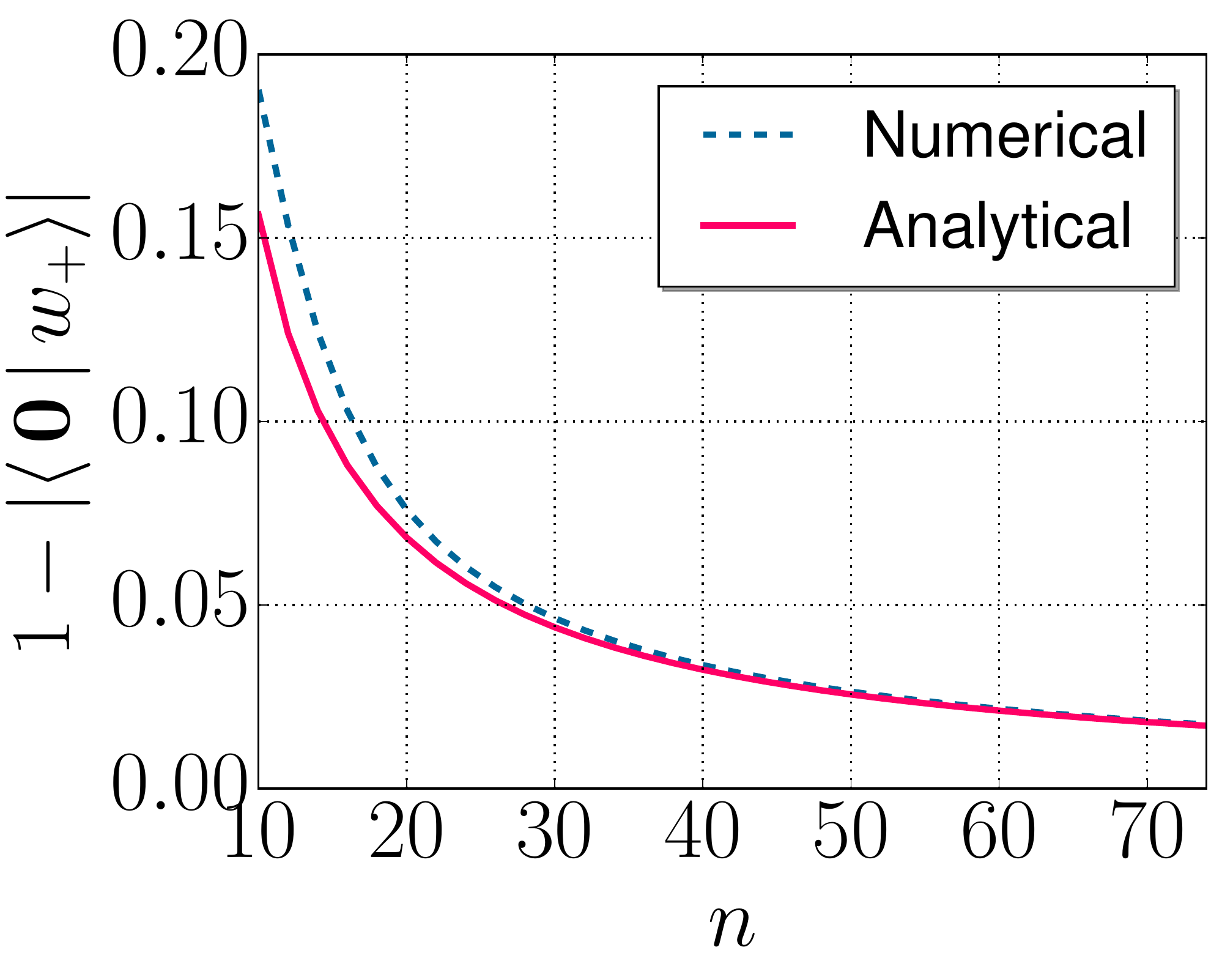}
\label{fig:w_plus}
}
\subfloat[]{\hspace{.8mm}
\includegraphics[width=.23\textwidth, height= 3.17cm]{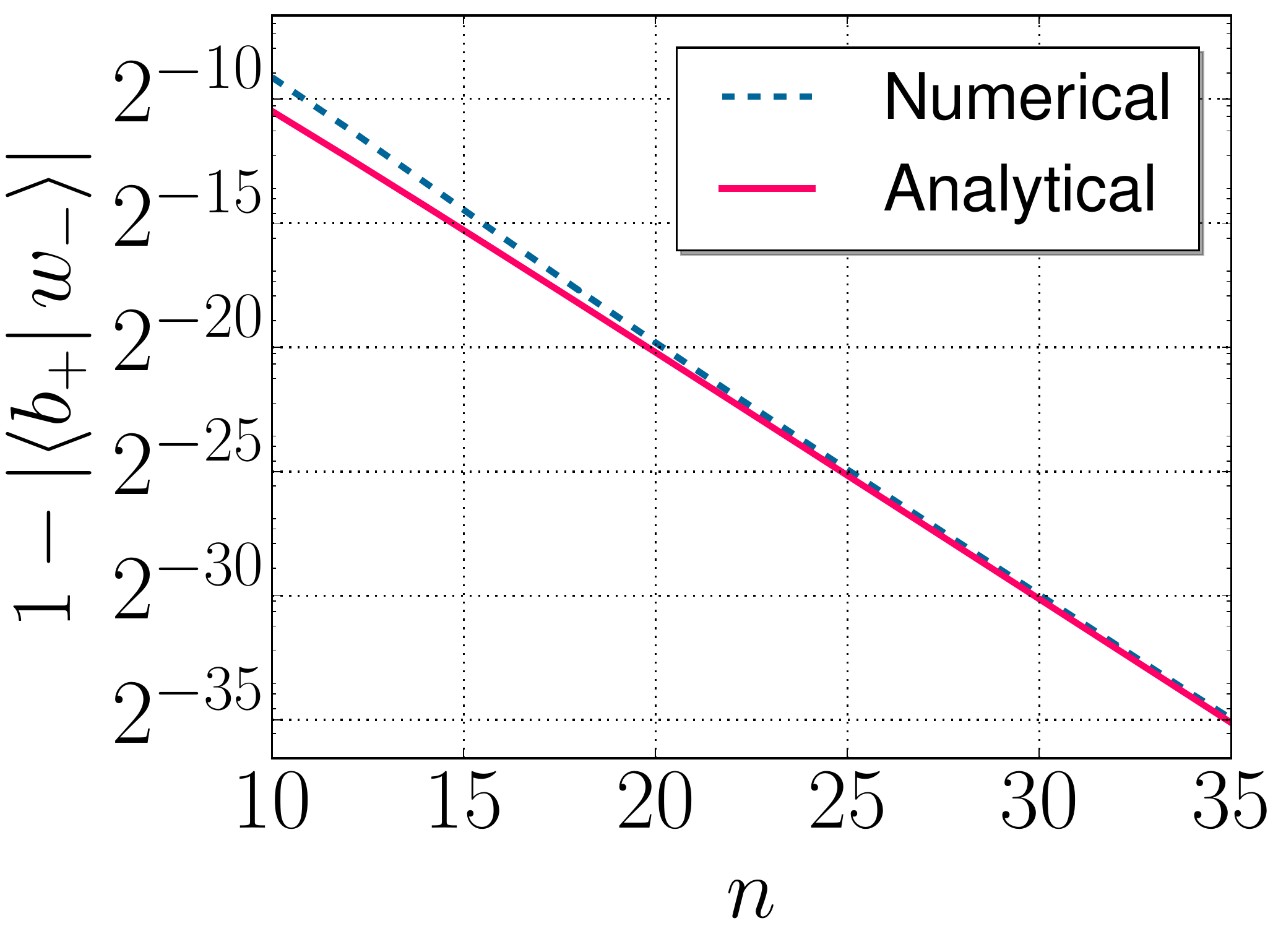}
\label{fig:psi_minus}
}
\caption{(a) Numerical and analytical (large-$n$ limit) results for the
infidelity $1-\lvert\braket{\bm 0}{\WW_+}\rvert$ as a function of the number of
qubits for $\gamma = \pi$, which vanishes polynomially as $n$ increases. The
numerical results are calculated by direct diagonalization of the matrix
$W(\pi)$ in the symmetric subspace; the analytical results use 
Eq.~(\ref{eq:fidelity_psi_+}), 
$\norm{\braket{\bm 0}{\WW_+}}\simeq (1-\pi^2/2n)^{1/4}$.
(b) Numerical and analytical (large-$n$
limit) results for the infidelity $1-\lvert\braket{\BB_+}{\WW_-}\rvert$ as a
function of the number of qubits $n$ for $\gamma = \pi$, which decreases
exponentially as $n$ increases. The numerical results come from direct
diagonalization, and the analytical results come from
Eq.~(\ref{eq:fidelity_psi_-_b}),
$\braket{\BB_+}{\WW_-} \simeq i\, \sqrt{2d/n}\, \big(1-\pi^2/2n\big)^{1/4}$.
} 
\label{fig:psi}
\end{figure}

\begin{figure}
\subfloat[]{\hspace{-2mm}
\includegraphics[width=.23\textwidth]{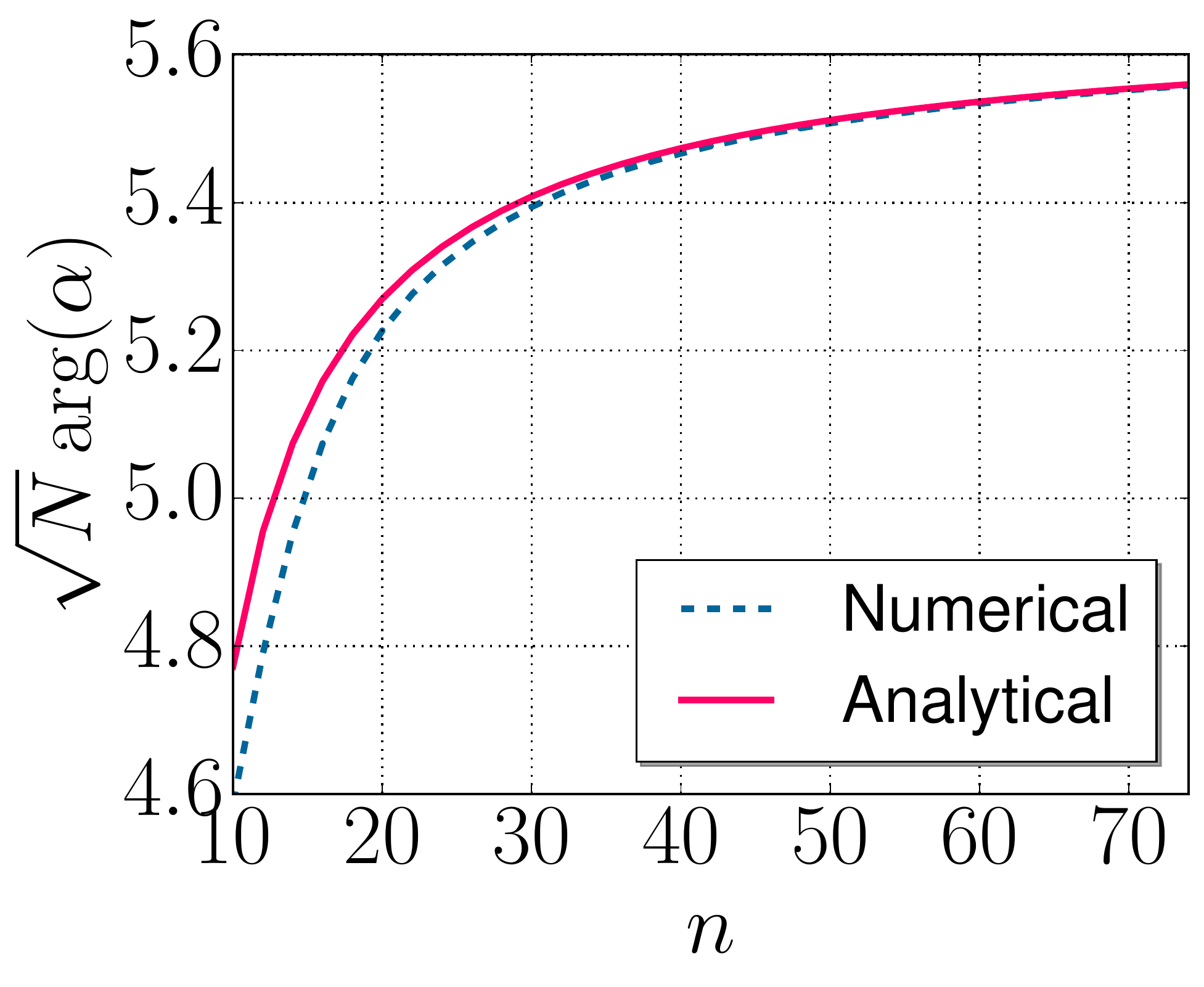}
\label{fig:arg_alpha}
}
\subfloat[]{\hspace{.4mm}
\includegraphics[width=.22\textwidth]{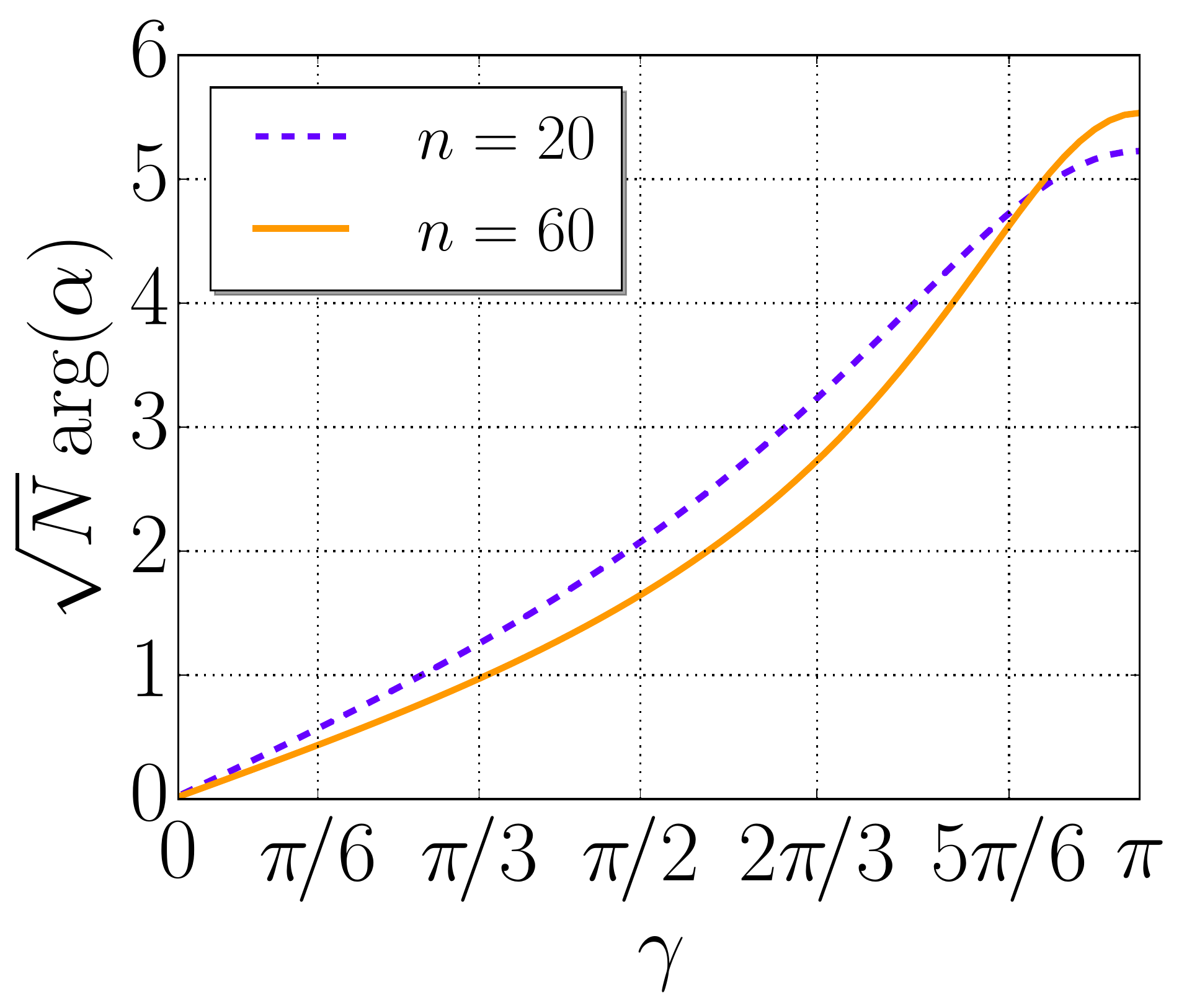}
\label{fig:arg_gamma}
}
\caption{(a) Numerical and analytical (large-$n$ limit) results for $\sqrt N
\arg(\alpha)$ as a function of $n$ for $\gamma = \pi$. (b) Numerical results
for $\sqrt N \arg(\alpha)$ as a function of $\gamma$. The numerical results
are computed using exact diagonalization. The analytical result comes
from Eq.~(\ref{eq:arg_modified}).
} 
\label{fig:arg}
\end{figure}

In Sec.~\ref{sec:eigen_V}, we derive approximate results for our algorithm in
the large-$n$ limit. For  $\gamma=\pi$, we find that 
$\norm{\braket{\bm 0}{\WW_+}}\simeq (1-\pi^2/2n)^{1/4}$ 
in Eq.~(\ref{eq:fidelity_psi_+}) (the
fidelity is smaller for $\gamma\neq \pi$). See Fig.~\subref*{fig:w_plus} for a
comparison of analytical and numerical results. We also find that
$\norm{\braket{\BB_+}{\WW_-}} \simeq 1- N^{-1}$ in
Eq.~(\ref{eq:fidelity_psi_-_c}). See Fig.~\subref*{fig:psi_minus} for a comparison
of analytical and numerical results.
Furthermore, we calculate that $\arg(\alpha) \simeq 4\sqrt{2}\, N^{-1/2}
(1-\pi^2/2n)^{1/4}$ in Eq.~(\ref{eq:arg_modified}). 
Figure~\subref*{fig:arg_alpha} shows a comparison of analytic and numerical results. 
Considering that the success probabilities of our
algorithm is about $1/2$ and each iteration $W(\gamma)$ calls the oracle twice,
the average query complexity of our algorithm is \begin{align}
 T(n) \simeq \frac{2\pi}{\arg(\alpha)}\simeq \frac{\pi}{2\sqrt 2}\, 2^{n/2}\,,
\end{align}
which differs from the optimal value presented in Ref.~\cite{zalka_grovers_1999} by a factor of $\sqrt 2$.

\section{An Intuitive Picture Using Spin-coherent states}
\label{sec:scs}

\begin{figure}
\centering
\subfloat[]{
\includegraphics[width=.228\textwidth]{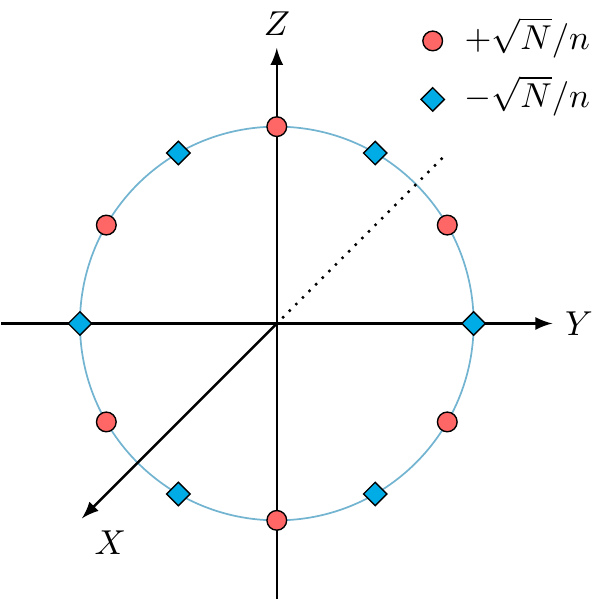}
\label{fig:b_plus}
}
\subfloat[]{
\includegraphics[width=.228\textwidth]{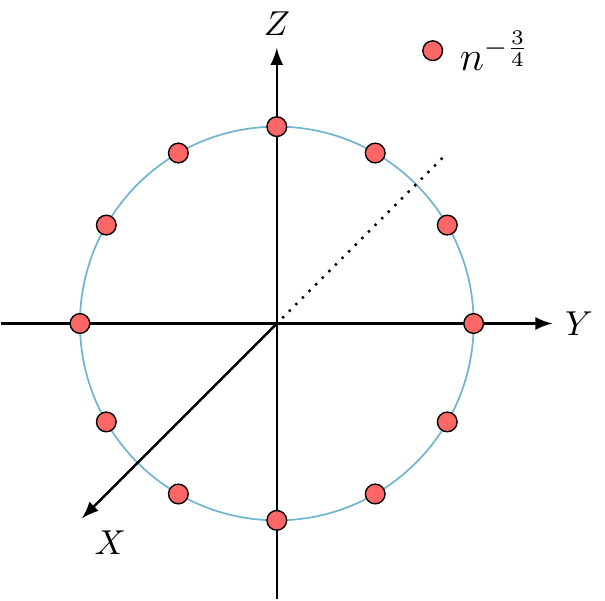}
\label{fig:b_zero}
}
\caption{Spin-coherent-state representation (a) for $\ket{\BB_+}$, where $\sqrt N/n$ is the order of the expansion coefficients in Eq.~(\ref{eq:b_+}), and (b) for $\ket{\BB_0}$, where $n^{-3/4}$ is the order of the expansion coefficients in Eq.~(\ref{eq:b_0}).}
\label{fig:b}
\end{figure}

This section gives the intuition behind our algorithm.
A representation using spin-coherent states~\cite{radcliffe_properties_1971,
perelomov_generalized_1986} is useful for understanding why the
algorithm works. Consider spin-coherent states of the form
\begin{align}\label{eq:spin_coherent}
 \ket{\psi(\theta)} = e^{-i\theta B/2}\ketb{\bm 0}\,,
\end{align}
where $\theta\in [\, 0, 2\pi\ssp)$; these states form an overcomplete basis for the symmetric subspace $\mathcal H_S$. We pay particular attention to a set of discrete angles $\theta_k = k\Delta \theta$, where $k=0,1,\ldots,n-1$ and $\Delta\theta = 2\pi/n$. Along with the dark state $\ket{\BB_-}$, this set of discrete spin-coherent states form a complete basis of $\mathcal H_S$. The state $\ket{\BB_+}$ can be expanded as [see Fig.~\ref{fig:b_plus}], 
\begin{align}\label{eq:b_+}
 \ket{\BB_+}
 &= \frac{1}{n\ssp \braket{\BB_+}{\bm 0}}\, \sum_{k=0}^{n-1} (-1)^k e^{-ik\pi B/n} \ket{\bm 0}\,,
\end{align}
where $\braket{\BB_+}{\bm 0} = \sqrt{2/N}$ is exponentially small in $n$. The normalization factor can be derived by noticing
\begin{align}
 \sum_{k=0}^{n-1} (-1)^k \bra{\BB_+} e^{-ik\pi B/n} \ket{\bm 0} = n \braket{\BB_+}{\bm 0}\,,
\end{align}
where we used the identity
\begin{align}\label{eq:rotation_a}
e^{-i\pi B/n} \ket{\BB_\pm} = - \ket{\BB_\pm} \,.
\end{align}
The expansion coefficient in Eq.~(\ref{eq:b_+}) can be derived by
noticing that $\ket{\BB_+}$ is orthogonal to any eigenstate of $B$ with an
eigenvalue other than $\pm n$. We will also need the eigenstate of $B$ with
eigenvalue $0$, \begin{align}
 \ket{\BB_0} \propto  P_S \Big(\ket{+}^{\otimes \frac{n}{2}}\otimes \ket{-}^{\otimes \frac{n}{2}}\Big)\,,
\end{align}
where $P_S$ is the projector onto $\mathcal H_S$. 
In other words, 
$\ket{\BB_0}$ is proportional to the sum of the $\binom{n}{n/2}$ terms that are
tensor products of the single-qubit states $\ket{+}$ and $\ket{-}$ with the
same number of occurrences, i.e., Hamming weight $n/2$ strings in the Hadamard
basis.
The overlap of this state with the target state is 
\begin{align}\label{eqn:b0approx}
 \norm{\braket{\BB_0}{\bm 0}}^2 &= \frac{n!}{(n/2)!\ssp (n/2)!} \,\frac{1}{2^n}
 \simeq \sqrt{\frac{ 2 }{\pi n}}\,,
\end{align}
which is only polynomially small. This state has the following expansion using the discrete spin-coherent states [see Fig.~\subref*{fig:b_zero}]:
\begin{align}\label{eq:b_0}
 \ket{\BB_0} = \frac{1}{n\ssp \braket{\BB_0}{\bm 0}}\, \sum_{k=0}^{n-1} e^{-ik\pi B/n} \ket{\bm 0}\,,
\end{align}
where $\braket{\BB_0}{\bm 0}$ is of order $n^{-1/4}$. It remains the same under the discrete rotation,
\begin{align}\label{eq:rotation_b}
e^{-i\pi B/n} \ket{\BB_0} = \ket{\BB_0} \,.
\end{align}

The unitary generated by the oracle takes the following form for $\gamma\ll 1$,
\begin{align}\label{eq:oracle_small}
 e^{-i\gamma C} \ket{\psi} &= \ket{\psi} + i\gamma\, \braket{\bm 0}{\psi} \ket{\bm 0} + \orderO(\gamma^2)\,,
\end{align}
where $\ket{\psi}$ is an arbitrary state. Putting Eqs.~(\ref{eq:W}), (\ref{eq:rotation_a}), (\ref{eq:rotation_b}), and (\ref{eq:oracle_small}) together, we have 
\begin{gather}
 W(\gamma)^{\frac{n}{2}} \ket{\BB_+} \simeq 
 \ket{\BB_+} + i \gamma\eta\ssp \ket{\BB_0}\,,\\[3pt]
 W(\gamma)^{\frac{n}{2}} \ket{\BB_0} \simeq 
 \ket{\BB_0} + i \gamma\eta\ssp \ket{\BB_+}\,,
\end{gather}
where 
\begin{align}
 \eta = n \,\braket{\BB_+}{\bm 0}\braket{\BB_0}{\bm 0} \simeq \sqrt[4]{2/\pi}\,  n^{3/4} N^{-1/2}\,.
\end{align}
Thus, the unitary $W(\gamma)^{n/2}$ approximately drives a transition between
$\ket{\BB_+}$ and $\ket{\BB_0}$ with the rate $\gamma\eta$. 
Applying the
unitary $W(\gamma)$ for order $n/\gamma\eta$ times, one can drive the state
$\ket{\BB_+}$ to a state close to $\ket{\BB_0}$. The probability of finding the
target state with $\ket{\BB_0}$ is only polynomially small in $n$ as opposed to
the exponentially small value with $\ket{\BB_+}$, achieving the quadratic
speedup in Grover's algorithm up to a logarithmic factor.

Although the case $\gamma\ll 1$ is illustrative, it requires logarithmically
many more calls to the oracle than Grover's original algorithm, and the 
probability of finding the target state is small.
Since $\eta$ is exponentially small in $n$, both 
$\ket{\BB_+}$ and $\ket{\BB_0}$ are 
close to eigenvectors for eigenvalues exponentially close to $1$.
This analysis suggests concentrating on the subspace spanned by 
$\{\ket{\WW_\alpha}, \ket{\WW_{\alpha^*}}\}$, where $\alpha$ and $\alpha^*$
are the eigenvalues closest to $1$. Indeed, we show in Sec.~\ref{sec:eigen_V}
that one can increase the success probability and reduce the number of
calls to the oracle by setting $\gamma=\pi$. In Fig.~\subref*{fig:arg_gamma},
$\arg(\alpha)$ is plotted as a function of $\gamma$. The reason behind why
$\gamma=\pi$ performs the best (or why it even works) seems unclear without a
tedious calculation. We give this
calculation in Sec.~\ref{sec:eigen_V}, after introducing a ``phase-space"
representation that will be useful in that analysis.

\section{Phase-space representations}
\label{sec:phase_space}

We introduce a  phase-space representation in this section which is
essential in the following section to the analytical solution of the success
probability and the query complexity of our algorithm. The phase-space
representation is based on the inner products of a quantum state with the spin-coherent states we introduced in Sec.~\ref{sec:scs}. 

Any state $\ket{\psi}\in \mathcal H_S$ can be uniquely determined by the inner
products $\brab{\bm 0}e^{i\theta B/2}\ketb{\psi}$. The $\chi$ function, 
\begin{align}
 \chi\big(\ket{\psi}, \theta\big) = \brab{\bm 0}e^{i\theta B/2}\ketb{\psi}\,,
\end{align}
fully determines the state $\ket{\psi}$ since the spin-coherent states $e^{-i\theta B/2}\,\ketb{\bm 0}$ for $\theta \in
[\,0,2\pi\ssp)$ are overcomplete for the symmetric subspace; the advantage of
this representation is that both $B$ and $C$ can be expressed concisely.
For even $n$, the $\chi$ function satisfies the periodic boundary condition
\begin{align}
\begin{split}
 \chi\big(\ket{\psi}, 2\pi\big) &= \brab{\bm 0}e^{i \pi B}\ketb{\psi}\\
 &= (-1)^n \braket{\bm 0}{\psi} = \chi\big(\ket{\psi},0\big)\,.
\end{split}
\end{align}
For the initial state in Eq.~(\ref{eq:initial}), we have
\begin{align}
 \chi\big(\ket{\psi_\mathrm{in}}, \theta\big) = \brab{\bm 0}e^{i\theta B/2}\ketb{\psi_\mathrm{in}} = \frac{e^{-i n\theta/2}}{\sqrt N}\,.
\end{align}
For the target state $\ket{\bm 0}$, we have
\begin{align}
 \chi\big(\ket{\bm 0}, \theta\big) = \brab{\bm 0}e^{i\theta B/2}\ketb{\bm 0}= \cos(\theta/2)^n\,.
\end{align}
The unitaries $e^{-i\phi B/2}$ and $e^{-i\gamma C}$ take simple forms,
\begin{align}\label{eq:rotaion_chi}
 &\chi\big(e^{-i\phi B/2}\ket{\psi}, \theta\big)   = \chi\big(\ket{\psi}, \theta -\phi\big)\,,\\[3pt]
\begin{split}
&\chi\big(e^{-i\gamma C}\ket{\psi}, \theta\big) \\
&\qquad = \chi\big(\ket{\psi}, \theta\big)+ (e^{i\gamma}-1) \, \chi\big(\ket{\psi}, 0\big)\cos(\theta/2)^n \,.
\end{split}\label{eq:oracle_chi}
\end{align}
For the discrete angles $\theta_k = 2 k \pi/n$, we introduce the notation
\begin{align}
 \chi_k\big(\ket{\psi}\big) = \brab{\bm 0}e^{ik \pi B/n}\ketb{\psi}\,.
\end{align}
The $\chi$ function of $\ket{\bm 0}$ will be used frequently, and we denote it as
\begin{align}\label{eq:xi}
 \xi_k \equiv \chi_k\big(\ket{\bm 0}\big) = \cos(k \pi/n)^n\,.
\end{align}
We will use the following identity repeatedly:
\begin{align}\label{eq:odd_sum}
 \sum_{k=0}^{n-1}(-1)^k\xi_k = n \braket{\bm 0}{\BB_+}^2= \frac{2 n}{N}\,.
\end{align}
For discrete angles, Eqs.~(\ref{eq:rotaion_chi}) and (\ref{eq:oracle_chi}) become
\begin{align}
 &\chi_k\big(e^{-i\pi B/n} \ket{\psi}\big) =  \chi_{k-1}\big(\ket{\psi}\big)\,,\label{eq:rotation_chi_dis}\\[3pt]
 &\chi_k\big(e^{-i\gamma C}\ket{\psi}\big) 
 = \chi_k\big(\ket{\psi}\big)+ (e^{i\gamma}-1) \chi_0\big(\ket{\psi}\big)\xi_k \,.\label{eq:oracle_chi_dis}
\end{align}
For the eigenstates of $B$ with eigenvalues $\pm n$, we have
\begin{align}\label{eq:b_n_chi}
 \chi_k\big(\ket{\BB_{n}}\big)= \chi_k\big(\ket{\BB_{-n}}\big)= (-1)^k N^{-1/2}\,,
\end{align}
where $\ket{\BB_{n}} = \ket{\psi_\mathrm{in}} = \ket{+}^{\otimes n}$ and $\ket{\BB_{-n}} = \ket{-}^{\otimes n}$. Since the discrete $\chi$ functions of $\ket{\BB_n}$ and $\ket{\BB_{-n}}$ are the same, it does not uniquely determine a state in the symmetric subspace with dimension $n+1$. The discrete $\chi$ function, however, is unique in the orthogonal space of $\ket{\BB_-} = \frac{1}{\sqrt 2} \big(\ket{\BB_{n}}-\ket{\BB_{-n}}\big)$. We will restrict our discussions into that subspace, and $\ket{\BB_-}$ is a dark state anyway. For $\ket{\BB_+} = \frac{1}{\sqrt 2} \big(\ket{\BB_{n}}+\ket{\BB_{-n}}\big)$, we have
\begin{align}\label{eq:b_+_chi}
 \chi_k\big(\ket{\BB_+}\big)= \sqrt 2\,(-1)^k N^{-1/2}\,.
\end{align}
The state $\ket{\BB_0}$ remains the same under $e^{-i\theta B/2}$, and its
$\chi$ function is a constant \begin{align}\label{eq:chi_b_0}
  \chi_k\big(\ket{\BB_0}\big)  = \chi_0\big(\ket{\BB_0}\big) \simeq \sqrt[4]{2 /\pi n}\,,
\end{align}
using the approximation in Eq.~(\ref{eqn:b0approx}).
To calculate the normalization factor of the $\chi$ representation, we need the
Fourier component 
\begin{align}
 \tilde\chi_j\big(\ket{\psi}\big) &= \frac{1}{n}\sum_{k=0}^{n-1} \chi_k\big(\ket{\psi}\big) \, e^{ijk\pi/n}\,,
\end{align}
where $j\in J \equiv \{-n,\ldots,-2,0, 2,\ldots, n\}$. The normalization condition is
\begin{align}\label{eq:normalization_chi}
 \norm{\braket{\psi}{\psi}}^2 &= \frac{N}{2}\,\normb{\tilde\chi_n\big(\ket{\psi}\big)}^2+ \sum_{j\in J'} \frac{\normb{\tilde\chi_j\big(\ket{\psi}\big)}^2}{\norm{\braket{\bm 0}{\BB_j}}^2}\,,
\end{align}
where $J'$ denotes the set $J\backslash\{\pm n\}$, and $\ket{\BB_j}$ is the eigenstate of $B$ whose eigenvalue is $j$, i.e. $B\ssp\ket{\BB_j} = j\ket{\BB_j}$.

\section{Analytical solutions}
\label{sec:eigen_V}

In this section, we solve the case $\gamma=\pi$ analytically using the phase
space representation introduced in Sec.~\ref{sec:phase_space}. Because $
e^{-i\pi C} = e^{i\pi C}$, it suffices to consider   \begin{align}
 V  &\equiv  e^{-i\pi B/n}\,e^{i\pi C}= \sqrt{W(\pi)}\,.
\end{align}
The state $\ket{\WW_\alpha}$, an eigenstate of $W$, is also an eigenstate of
$V$. The new eigenvalue is $\beta = \alpha^{1/2}$ ($\beta$ is close to $-1$).
The remainder of this section is devoted to finding the relevant eigenvalues
and eigenstates of $V$. The eigenvalues determine the query complexity of our
algorithm, while the corresponding eigenvectors determine the probability of
success. 

We introduce the unnormalized $\chi$ functions of the eigenstate
$\ket{\WW_\alpha}$, \begin{align}\label{eq:unnormalized_chi}
 \xx_k \equiv \chi_k(\ket{\WW_\alpha})\big/\chi_0(\ket{\WW_\alpha})\,,
\end{align}
which satisfies $\xx_0=1$. Using Eqs.~(\ref{eq:rotation_chi_dis}) and (\ref{eq:oracle_chi_dis}), we have 
\begin{align}\label{eq:unnormalized_eigenstate}
 \xx_k = \beta^k + 2\sum_{\ell=1}^k \beta^{k-\ell} \xi_\ell\,,
\end{align}
where $\xi_\ell$ is defined in Eq.~(\ref{eq:xi}). The periodic boundary condition $\xx_n = \xx_0$ gives the eigenvalue equation for $\beta$,
\begin{align}\label{eq:eign_val_eq_a}
 ( 1+ \beta^{n})/2+\beta^{n-1}\xi_1+ \cdots +\beta^2\xi_{n-2}+\beta\xi_{n-1} = 0\,.
\end{align}
Because Eq.~(\ref{eq:eign_val_eq_a}) contains only real coefficients, $\beta^*$ is also a solution to it [it comes from the symmetry~(\ref{eq:time_reversal})]. For $\beta\simeq -1$, we have
\begin{align}\label{eq:eign_val}
  \beta = -\sqrt{1-\delta^2} - i\delta\simeq -1 - i\delta + \delta^2/2,
\end{align}
where $\delta > 0$ is a small real parameter of order $1/\sqrt N$. Putting
Eq.~(\ref{eq:eign_val}) into Eq.~(\ref{eq:eign_val_eq_a}) and keeping only terms up to order $\delta^2$, we have \begin{widetext}
\begin{align}\label{eq:eign_val_eq_b}
   0 &= 1 + in\delta/2-n^2\delta^2/4+\sum_{k=1}^{n-1}(-1)^k\Big(1+(n-k) \big[i\delta-(n-k)\delta^2/2)\big]\Big)\xi_k + \orderO(\delta^3)\nonumber\\
   &=\sum_{k=0}^{n-1}(-1)^k \xi_k +i\delta \bigg( \frac{n}{2} + \sum_{k=1}^{n-1}(-1)^k (n-k)\, \xi_k \bigg)-\frac{\delta^2}{2}\bigg( \frac{n^2}{2}+ \sum_{k=1}^{n-1}(-1)^k (n-k)^2\, \xi_k\bigg)+\orderO(\delta^3)\,. 
\end{align}
\end{widetext}
The coefficient of the term with $i\delta$ in Eq.~(\ref{eq:eign_val_eq_b}) is
\begin{align}\label{eq:imaginary_sum}
   \frac{n}{2}  + \sum_{k=1}^{n-1}(-1)^k (n-k)\, \xi_k 
  &= \frac{n}{2} \sum_{k=0}^{n-1}(-1)^k \xi_k = \frac{n^2}{N}\,,  
\end{align}
where we have used Eq.~(\ref{eq:odd_sum}); therefore, the pure imaginary term in Eq.~(\ref{eq:eign_val_eq_b}) is of order $\delta^3$ and can be neglected. Comparing the real
parts at both sides of Eq.~(\ref{eq:eign_val_eq_b}), we have
\begin{align}\label{eq:delta2}
  \delta^2 \simeq \frac{2\sum_{k=0}^{n-1}(-1)^k \xi_k}{n^2/2+\sum_{k=1}^{n-1}(-1)^k (n-k)^2\, \xi_k}\,.
\end{align}
While the numerator in Eq.~(\ref{eq:delta2}) has already been solved in
Eq.~(\ref{eq:odd_sum}), the denominator is harder to calculate. We write the
denominator as \begin{align}\label{eq:d}
 d = n^2/2+\sum_{k=1}^{n-1}(-1)^k (n-k)^2\, \xi_k\,,
\end{align}
and we will solve it later (but remember $d\sim n$). Putting Eqs.~(\ref{eq:odd_sum}) and (\ref{eq:d}) into Eq.~(\ref{eq:delta2}), we have the formal solution
\begin{align}\label{eq:delta_d}
 \delta = 2\sqrt{n/d}\, N^{-1/2}\,,
\end{align}
where $d$ is to be determined. 

Let $\xx_k^+ $ and $\xx_k^-$ be the real and imaginary parts of the
function $\xx_k$ defined in Eq.~(\ref{eq:unnormalized_chi}); we have
\begin{align}
 &\xx_k^+ = \chi_k(\ket{\WW_+})\big/\chi_0(\ket{\WW_+})\,,\\
 &\xx_k^- = \chi_k(\ket{\WW_-})\big/\chi_0(\ket{\WW_+})\,,
\end{align}
where we use the identity $\chi_0(\ket{\WW_+}) = \sqrt 2 \, \chi_0(\ket{\WW_\alpha})$. 
The normalization factor
$\chi_0(\ket{\WW_+})=\norm{\braket{\bm 0}{\WW_+}}$ determines the overlap and can be calculated by
using Eq.~(\ref{eq:normalization_chi}). Separating the real and imaginary parts
in the expansion~(\ref{eq:unnormalized_eigenstate}), we have \begin{align}
 &\xx_k^+ \simeq (-1)^k + 2\sum_{\ell=1}^k (-1)^{k-\ell} \xi_\ell\,,\label{eq:xx_real}\\[-2pt]
 &\xx_k^- \simeq i\delta \Big((-1)^k  k  + 2\sum_{\ell=1}^k (-1)^{k-\ell}  (k-\ell) \xi_\ell\Big)\,,\label{eq:xx_imag}
\end{align}
where higher-order terms in $\delta$ are neglected. The $j$th Fourier component of $\xx^+$ is
\begin{align}
 \tilde\xx_j^+ &=  \frac{2}{n(1+e^{ij\pi/n})} \sum_{k=0}^{n-1} \xi_k \Big(e^{ijk\pi/n} - (-1)^k\Big)\nonumber\\
 &\simeq \frac{2}{1+e^{ij\pi/n}}\, \norm{\braket{\bm 0}{\BB_j}}^2 \,,
\end{align}
where $j \in J\equiv \{-n,\ldots,-2,0, 2,\ldots, n\}$. Using the normalization condition~(\ref{eq:normalization_chi}), we have
\begin{align}\label{eq:fidelity_psi_+_a}
 \frac{1}{\norm{\braket{\bm 0}{\WW_+}}^2}&\simeq \sum_{j\in J'} \frac{\norm{\tilde\xx_j^+}^2}{ \norm{\braket{\bm 0}{\BB_j}}^2}\simeq \sum_{j\in J'} \frac{2\, \norm{\braket{\bm 0}{\BB_j}}^2}{1+\cos(j\pi/n)}\,,
\end{align}
where $J'=J\backslash\{\pm n\}$ and the exponentially small term proportional to $\norm{\tilde\xx_n^+}^2$ is neglected. For $\norm{j}\ll n$, we have
\begin{align}\label{eq:modify}
\frac{2}{1+\cos(j\pi/n)} \simeq 1 + \pi^2\tau^2\simeq e^{\pi^2\tau^2}\,,
\end{align}
where $\tau \equiv j/2n$. The squared fidelity $\norm{\braket{\bm 0}{\BB_j}}^2$ can also be approximated by a Gaussian for $\tau\ll 1$, 
\begin{align}\label{eq:Gaussian}
 \norm{\braket{\bm 0}{\BB_j}}^2 &= \frac{n!}{n_+!\,n_-!} \,\frac{1}{2^n}\simeq \frac{2\ssp e^{-2 n\tau^2}}{\sqrt{2\pi n}} \,,
\end{align}
where $n_\pm = (n\pm j)/2 = n(1/2\pm\tau)$. The term in Eq.~(\ref{eq:modify}) modifies the variance of the Gaussian~(\ref{eq:Gaussian}) by a factor of $2n/(2n-\pi^2)$, and thus we have
\begin{align}\label{eq:ratio_variance}
\sum_j \frac{2\, \norm{\braket{\bm 0}{\BB_j}}^2}{1+\cos(j\pi/n)}\simeq \sqrt{\frac{2n}{2n-\pi^2}}\,,
\end{align}
where we used the condition $\sum_{j\in J'} \,\norm{\braket{\bm 0}{\BB_j}}^2 \simeq 1$. 
Putting Eq.~(\ref{eq:ratio_variance}) into Eq.~(\ref{eq:fidelity_psi_+_a}), we have 
\begin{align}\label{eq:fidelity_psi_+}
 \norm{\braket{\bm 0}{\WW_+}}&\simeq \big(1-\pi^2/2n\big)^{1/4}\,,
\end{align}
which becomes arbitrarily close to $1$ for large $n$; see Fig.~\subref*{fig:w_plus} for a comparison to numerics. 

To derive the fidelity $\lvert\braket{\BB_+}{\WW_-}\rvert$,  we notice
\begin{align}\label{eq:sum_relation}
 \xx_k^- + \xx_{k+1}^- = - i\delta \xx_k^+\,,
\end{align}
which is proportional to the $\chi$ function of $\ket{\WW_+}$. Because $\ket{\WW_+}\simeq e^{-i\pi B/n}\ket{\WW_+}$, Eq.~(\ref{eq:sum_relation}) implies that 
\begin{align}\label{eq:w_-_decompo}
 \ket{\WW_-} \simeq  \braket{\BB_+}{\WW_-}\ket{\BB_+} -\frac{i\delta}{2}\, \ket{\WW_+}\,.
\end{align}
Thus, we can estimate the fidelity
\begin{align}\label{eq:fidelity_psi_-_a}
 \norm{\braket{\BB_+}{\WW_-}} \simeq 1- \delta^2/8\,,
\end{align}
which is exponentially close to $1$ ($\delta^2\sim N^{-1}$). 

The value of $\delta$, however, is only formally solved in Eq.~(\ref{eq:delta_d}). We still need to determine the value of $d$ defined in Eq.~(\ref{eq:d}). The Fourier component of $\xx^-$ corresponding to $\ket{\BB_+}$ is
\begin{widetext}
\begin{align}
 \frac{1}{n}\sum_{k=0}^{n-1} (-1)^k \xx_k^- &= \frac{i\delta}{n} \sum_{k=0}^{n-1} (-1)^k\Big((-1)^k k  + 2\sum_{\ell=1}^k (-1)^{k-\ell}  (k-\ell) \xi_\ell\Big)\nonumber\\
 &= \frac{i\delta}{n} \Big(\half n (n-1)  + 2\sum_{\ell=1}^{n-1} \sum_{k=\ell}^{n-1} (-1)^\ell  (k-\ell) \xi_\ell\Big)\nonumber\\
 &= \frac{i\delta}{n} \Big(\frac{n^2}{2} +\sum_{\ell=1}^{n-1} (-1)^\ell  (n-\ell)^2 \xi_\ell - \frac{n}{2}  - \sum_{\ell=1}^{n-1} (-1)^\ell  (n-\ell) \xi_\ell\Big) = i\delta \big(d/n - n/N\big)\,,\label{eq:xx_-_B_+}
\end{align}
\end{widetext}
where we used Eqs.~(\ref{eq:imaginary_sum}) and (\ref{eq:d}) in the last step. 
By neglecting the higher order term in Eq.~(\ref{eq:xx_-_B_+}), we have
\begin{align}\label{eq:fidelity_psi_-_b}
\begin{split}
 \braket{\BB_+}{\WW_-} &\simeq  i\delta (d/n)\sqrt{N/2}\: \norm{\braket{\bm 0}{\WW_+}}\\[3pt]
 &\simeq i\, \sqrt{2d/n}\, \big(1-\pi^2/2n\big)^{1/4}\,.
\end{split}
\end{align}
where we used Eqs.~(\ref{eq:delta_d}) and (\ref{eq:fidelity_psi_+}). Comparing Eq.~(\ref{eq:fidelity_psi_-_b}) with Eq.~(\ref{eq:fidelity_psi_-_a}), we have
\begin{align}
 d\simeq \frac{n}{2}\big(1-\pi^2/2n\big)^{-1/2}\,.
\end{align}
Putting this result into Eq.~(\ref{eq:delta_d}), we have
\begin{align}\label{eq:delta}
 \delta \simeq 2\sqrt{2}\, N^{-1/2} \big(1-\pi^2/2n\big)^{1/4}\,.
\end{align}
The argument of $\alpha$ thus takes the form
\begin{align}\label{eq:arg_modified}
 \arg(\alpha) \simeq 2\delta \simeq 4\sqrt{2}\, N^{-1/2} \big(1-\pi^2/2n\big)^{1/4}\,,
\end{align}
which conforms with the numerical result in Fig.~\subref*{fig:arg_alpha}. Putting
Eq.~(\ref{eq:delta}) into Eq.~(\ref{eq:fidelity_psi_-_a}), we have the fidelity
\begin{align}\label{eq:fidelity_psi_-_c}
 \norm{\braket{\BB_+}{\WW_-}} \simeq 1 - N^{-1}\,,
\end{align}
where we drop the factor $(1-\pi^2/2n)^{1/2}$, because it is of the same order as the approximation made in Eq.~(\ref{eq:w_-_decompo}). 

We calculate the fidelities $\norm{\braket{\bm 0}{\WW_+}}$ and $\lvert\braket{\BB_+}{\WW_-}\rvert$ numerically for $\gamma\neq \pi$ and find that they are always less than the corresponding values at $\gamma=\pi$. The alternating signs in $e^{i\gamma C}$ and $e^{-i\gamma C}$ are important for $\gamma\neq \pi$; the probability of finding the target state almost vanishes when the same sign is used (localized eigenstates). 

\section{Check the solution}
\label{sec:check}

Because the success probability of our algorithm is about $1/2$, it may not be
very efficient to use a majority vote approach to find the marked bit string
with high probability. Here, we describe a method to check whether the marked
bit string has been found systematically. 

Suppose that we have found the bit string $\ket{\bms}$ at the output of the
circuit. Apply a $\pi/2$ pulse on an arbitrary qubit, creating an even
superposition of the bit string $\ket{\bms}$ and a flipped bit sting
$\ket{\bms'}$. Then apply the unitary $e^{i\pi C}$ to the system; this
step flips the sign of the target bit string. Finally, apply a $-\pi/2$
pulse to the selected qubit and measure in the computational basis. One of the
two bit stings $\ket{\bms}$ and $\ket{\bms'}$ must be the target if the
measurement outcome is the bit sting $\ket{\bms'}$; otherwise, neither of the
two bit strings is the target. To distinguish whether the bit string
$\ket{\bms}$ or $\ket{\bms'}$ is the target, do the whole
procedure over again on a different qubit.

\section{Conclusion}
\label{sec:conclusion}

Inspired by the QAOA proposed by Farhi \ea~\cite{farhi_quantum_2014,
farhi_quantum_2014-1}, we presented a circuit-based quantum algorithm to search for a needle in a
haystack. We showed that Grover's diffusion operator can be replaced by the
transverse field, which requires only single-qubit gates, without 
sacrificing the
quadratic quantum speedup. As single-qubit gates can usually be carried out
much more efficiently than multi-qubit gates in practice, our algorithm offers
a mild implementation advantage for Grover's unstructured search and its
variants. 
This circuit model approach 
can take advantage of fault-tolerant
error-correcting schemes; it is not known how, and could be 
impossible, to achieve fault tolerance in a purely adiabatic
model~\cite{young_error_2013}. 

We construct a simple periodic sequence of gates that
induces a closed transition between two states which have large overlaps with
the initial and target states, respectively. The query complexity of our
algorithm is $T(n) \simeq (\pi/2\sqrt 2\,)\, 2^{n/2}$, differing from the
optimal value proved in~\cite{zalka_grovers_1999} by only a constant factor of
$\sqrt
2$. 
Our algorithm provides a QAOA circuit that exhibits
a quantum advantage at an intermediate number of iterations $p$, 
$p \gg 1$, and the algorithm is not derived from Trotterization of an 
AQO algorithm, demonstrating the breadth of the QAOA framework.
It remains an open question whether QAOA circuits provide a quantum
advantage for approximate optimization.

It is generally hard to find the optimal parameters in the QAOA when the number
of iterations of the algorithm is large. Our work demonstrates that even simple
periodic dynamics generated by the transverse field and the problem Hamiltonian
can induce interesting transitions between a problem-independent state and an
approximate target state. It offers a strategy to drastically simplify the
optimization of the parameters in QAOA by restricting them to be periodic. For
Grover's unstructured search, such simplification yields a near-optimal
solution
to the problem. It will be interesting to see how well this strategy works 
for more general cases. 

Our algorithm can be understood intuitively using a spin-coherent-state
representation, where the weights of the basis states evolve in a simple way
under the unitaries generated by the driver and the oracle. We also use a 
phase-space representation based on spin-coherent states to analyze the
composite unitary in our algorithm. 
The eigenstates (up to normalization
factors) of the composite unitary take explicit forms in this representation,
and the eigenvalue equation can be readily derived using the periodic boundary
condition. This enables us to solve the eigenstates and eigenvalues to exponential
precision in $n$. It is worth exploring the extent to which such a 
representation is effective for more general quantum heuristic algorithms.

\begin{acknowledgments}
The authors thank Salvatore Mandr\`{a} and Davide Venturelli for enlightening and helpful
discussions. The authors would like to acknowledge support from the NASA
Advanced Exploration Systems program and NASA Ames Research Center. This work
was also supported in part by the AFRL Information Directorate under Grant No.
F4HBKC4162G001 and the Office of the Director of National Intelligence (ODNI). The
views and conclusions contained herein are those of the authors and should not
be interpreted as necessarily representing the official policies or
endorsements, either expressed or implied, of ODNI, AFRL, or the U.S.
Government.  The U.S. Government is authorized to reproduce and distribute
reprints for Governmental purpose notwithstanding any copyright annotation
thereon.
\end{acknowledgments}


\begin{thebibliography}{19}%
\makeatletter
\providecommand \@ifxundefined [1]{%
 \@ifx{#1\undefined}
}%
\providecommand \@ifnum [1]{%
 \ifnum #1\expandafter \@firstoftwo
 \else \expandafter \@secondoftwo
 \fi
}%
\providecommand \@ifx [1]{%
 \ifx #1\expandafter \@firstoftwo
 \else \expandafter \@secondoftwo
 \fi
}%
\providecommand \natexlab [1]{#1}%
\providecommand \enquote  [1]{``#1''}%
\providecommand \bibnamefont  [1]{#1}%
\providecommand \bibfnamefont [1]{#1}%
\providecommand \citenamefont [1]{#1}%
\providecommand \href@noop [0]{\@secondoftwo}%
\providecommand \href [0]{\begingroup \@sanitize@url \@href}%
\providecommand \@href[1]{\@@startlink{#1}\@@href}%
\providecommand \@@href[1]{\endgroup#1\@@endlink}%
\providecommand \@sanitize@url [0]{\catcode `\\12\catcode `\$12\catcode
  `\&12\catcode `\#12\catcode `\^12\catcode `\_12\catcode `\%12\relax}%
\providecommand \@@startlink[1]{}%
\providecommand \@@endlink[0]{}%
\providecommand \url  [0]{\begingroup\@sanitize@url \@url }%
\providecommand \@url [1]{\endgroup\@href {#1}{\urlprefix }}%
\providecommand \urlprefix  [0]{URL }%
\providecommand \Eprint [0]{\href }%
\providecommand \doibase [0]{http://dx.doi.org/}%
\providecommand \selectlanguage [0]{\@gobble}%
\providecommand \bibinfo  [0]{\@secondoftwo}%
\providecommand \bibfield  [0]{\@secondoftwo}%
\providecommand \translation [1]{[#1]}%
\providecommand \BibitemOpen [0]{}%
\providecommand \bibitemStop [0]{}%
\providecommand \bibitemNoStop [0]{.\EOS\space}%
\providecommand \EOS [0]{\spacefactor3000\relax}%
\providecommand \BibitemShut  [1]{\csname bibitem#1\endcsname}%
\let\auto@bib@innerbib\@empty
\bibitem [{\citenamefont {Farhi}\ \emph
  {et~al.}(2014{\natexlab{a}})\citenamefont {Farhi}, \citenamefont
  {Goldstone},\ and\ \citenamefont {Gutmann}}]{farhi_quantum_2014}%
  \BibitemOpen
  \bibfield  {author} {\bibinfo {author} {\bibfnamefont {Edward}\ \bibnamefont
  {Farhi}}, \bibinfo {author} {\bibfnamefont {Jeffrey}\ \bibnamefont
  {Goldstone}}, \ and\ \bibinfo {author} {\bibfnamefont {Sam}\ \bibnamefont
  {Gutmann}},\ }\bibfield  {title} {\enquote {\bibinfo {title} {A {Quantum}
  {Approximate} {Optimization} {Algorithm}},}\ }\href
  {http://arxiv.org/abs/1411.4028} {\bibfield  {journal} {\bibinfo  {journal}
  {arXiv:1411.4028}\ } (\bibinfo {year} {2014}{\natexlab{a}})}\BibitemShut
  {NoStop}%
\bibitem [{\citenamefont {Farhi}\ \emph
  {et~al.}(2014{\natexlab{b}})\citenamefont {Farhi}, \citenamefont
  {Goldstone},\ and\ \citenamefont {Gutmann}}]{farhi_quantum_2014-1}%
  \BibitemOpen
  \bibfield  {author} {\bibinfo {author} {\bibfnamefont {Edward}\ \bibnamefont
  {Farhi}}, \bibinfo {author} {\bibfnamefont {Jeffrey}\ \bibnamefont
  {Goldstone}}, \ and\ \bibinfo {author} {\bibfnamefont {Sam}\ \bibnamefont
  {Gutmann}},\ }\bibfield  {title} {\enquote {\bibinfo {title} {A {Quantum}
  {Approximate} {Optimization} {Algorithm} {Applied} to a {Bounded}
  {Occurrence} {Constraint} {Problem}},}\ }\href
  {http://arxiv.org/abs/1412.6062} {\bibfield  {journal} {\bibinfo  {journal}
  {arXiv:1412.6062}\ } (\bibinfo {year} {2014}{\natexlab{b}})}\BibitemShut
  {NoStop}%
\bibitem [{\citenamefont {Grover}(1996)}]{grover_fast_1996}%
  \BibitemOpen
  \bibfield  {author} {\bibinfo {author} {\bibfnamefont {Lov~K.}\ \bibnamefont
  {Grover}},\ }\bibfield  {title} {\enquote {\bibinfo {title} {A {Fast}
  {Quantum} {Mechanical} {Algorithm} for {Database} {Search}},}\ }in\ \href
  {\doibase 10.1145/237814.237866} {\emph {\bibinfo {booktitle} {Proceedings of
  the {Twenty}-eighth {Annual} {ACM} {Symposium} on {Theory} of
  {Computing}}}},\ \bibinfo {series and number} {{STOC} '96}\ (\bibinfo
  {publisher} {ACM},\ \bibinfo {address} {New York, NY, USA},\ \bibinfo {year}
  {1996})\ pp.\ \bibinfo {pages} {212--219}\BibitemShut {NoStop}%
\bibitem [{\citenamefont {Goldman}\ and\ \citenamefont
  {Dalibard}(2014)}]{goldman_periodically_2014}%
  \BibitemOpen
  \bibfield  {author} {\bibinfo {author} {\bibfnamefont {N.}~\bibnamefont
  {Goldman}}\ and\ \bibinfo {author} {\bibfnamefont {J.}~\bibnamefont
  {Dalibard}},\ }\bibfield  {title} {\enquote {\bibinfo {title} {Periodically
  {Driven} {Quantum} {Systems}: {Effective} {Hamiltonians} and {Engineered}
  {Gauge} {Fields}},}\ }\href {\doibase 10.1103/PhysRevX.4.031027} {\bibfield
  {journal} {\bibinfo  {journal} {Physical Review X}\ }\textbf {\bibinfo
  {volume} {4}},\ \bibinfo {pages} {031027} (\bibinfo {year}
  {2014})}\BibitemShut {NoStop}%
\bibitem [{\citenamefont {Wecker}\ \emph {et~al.}(2016)\citenamefont {Wecker},
  \citenamefont {Hastings},\ and\ \citenamefont {Troyer}}]{wecker2016training}%
  \BibitemOpen
  \bibfield  {author} {\bibinfo {author} {\bibfnamefont {Dave}\ \bibnamefont
  {Wecker}}, \bibinfo {author} {\bibfnamefont {Matthew~B}\ \bibnamefont
  {Hastings}}, \ and\ \bibinfo {author} {\bibfnamefont {Matthias}\ \bibnamefont
  {Troyer}},\ }\bibfield  {title} {\enquote {\bibinfo {title} {Training a
  quantum optimizer},}\ }\href@noop {} {\bibfield  {journal} {\bibinfo
  {journal} {Physical Review A}\ }\textbf {\bibinfo {volume} {94}},\ \bibinfo
  {pages} {022309} (\bibinfo {year} {2016})}\BibitemShut {NoStop}%
\bibitem [{\citenamefont {Farhi}\ and\ \citenamefont
  {Harrow}(2016)}]{farhi_quantum_2016}%
  \BibitemOpen
  \bibfield  {author} {\bibinfo {author} {\bibfnamefont {Edward}\ \bibnamefont
  {Farhi}}\ and\ \bibinfo {author} {\bibfnamefont {Aram~W.}\ \bibnamefont
  {Harrow}},\ }\bibfield  {title} {\enquote {\bibinfo {title} {Quantum
  {Supremacy} through the {Quantum} {Approximate} {Optimization}
  {Algorithm}},}\ }\href {http://arxiv.org/abs/1602.07674} {\bibfield
  {journal} {\bibinfo  {journal} {arXiv:1602.07674}\ } (\bibinfo {year}
  {2016})}\BibitemShut {NoStop}%
\bibitem [{\citenamefont {Preskill}(2012)}]{preskill_quantum_2012}%
  \BibitemOpen
  \bibfield  {author} {\bibinfo {author} {\bibfnamefont {John}\ \bibnamefont
  {Preskill}},\ }\bibfield  {title} {\enquote {\bibinfo {title} {Quantum
  computing and the entanglement frontier},}\ }\href
  {http://arxiv.org/abs/1203.5813} {\bibfield  {journal} {\bibinfo  {journal}
  {arXiv:1203.5813}\ } (\bibinfo {year} {2012})}\BibitemShut {NoStop}%
\bibitem [{\citenamefont {Boixo}\ \emph {et~al.}(2016)\citenamefont {Boixo},
  \citenamefont {Isakov}, \citenamefont {Smelyanskiy}, \citenamefont {Babbush},
  \citenamefont {Ding}, \citenamefont {Jiang}, \citenamefont {Martinis},\ and\
  \citenamefont {Neven}}]{boixo_characterizing_2016}%
  \BibitemOpen
  \bibfield  {author} {\bibinfo {author} {\bibfnamefont {Sergio}\ \bibnamefont
  {Boixo}}, \bibinfo {author} {\bibfnamefont {Sergei~V.}\ \bibnamefont
  {Isakov}}, \bibinfo {author} {\bibfnamefont {Vadim~N.}\ \bibnamefont
  {Smelyanskiy}}, \bibinfo {author} {\bibfnamefont {Ryan}\ \bibnamefont
  {Babbush}}, \bibinfo {author} {\bibfnamefont {Nan}\ \bibnamefont {Ding}},
  \bibinfo {author} {\bibfnamefont {Zhang}\ \bibnamefont {Jiang}}, \bibinfo
  {author} {\bibfnamefont {John~M.}\ \bibnamefont {Martinis}}, \ and\ \bibinfo
  {author} {\bibfnamefont {Hartmut}\ \bibnamefont {Neven}},\ }\bibfield
  {title} {\enquote {\bibinfo {title} {Characterizing {Quantum} {Supremacy} in
  {Near}-{Term} {Devices}},}\ }\href {http://arxiv.org/abs/1608.00263}
  {\bibfield  {journal} {\bibinfo  {journal} {arXiv:1608.00263}\ } (\bibinfo
  {year} {2016})}\BibitemShut {NoStop}%
\bibitem [{\citenamefont {Barak}\ \emph {et~al.}(2015)\citenamefont {Barak},
  \citenamefont {Moitra}, \citenamefont {O'Donnell}, \citenamefont
  {Raghavendra}, \citenamefont {Regev}, \citenamefont {Steurer}, \citenamefont
  {Trevisan}, \citenamefont {Vijayaraghavan}, \citenamefont {Witmer},\ and\
  \citenamefont {Wright}}]{barak_beating_2015_published}%
  \BibitemOpen
  \bibfield  {author} {\bibinfo {author} {\bibfnamefont {Boaz}\ \bibnamefont
  {Barak}}, \bibinfo {author} {\bibfnamefont {Ankur}\ \bibnamefont {Moitra}},
  \bibinfo {author} {\bibfnamefont {Ryan}\ \bibnamefont {O'Donnell}}, \bibinfo
  {author} {\bibfnamefont {Prasad}\ \bibnamefont {Raghavendra}}, \bibinfo
  {author} {\bibfnamefont {Oded}\ \bibnamefont {Regev}}, \bibinfo {author}
  {\bibfnamefont {David}\ \bibnamefont {Steurer}}, \bibinfo {author}
  {\bibfnamefont {Luca}\ \bibnamefont {Trevisan}}, \bibinfo {author}
  {\bibfnamefont {Aravindan}\ \bibnamefont {Vijayaraghavan}}, \bibinfo {author}
  {\bibfnamefont {David}\ \bibnamefont {Witmer}}, \ and\ \bibinfo {author}
  {\bibfnamefont {John}\ \bibnamefont {Wright}},\ }\bibfield  {title} {\enquote
  {\bibinfo {title} {Beating the random assignment on constraint satisfaction
  problems of bounded degree},}\ }in\ \href
  {https://eccc.weizmann.ac.il/report/2015/082/} {\emph {\bibinfo {booktitle}
  {APPROX-RANDOM}}}\ (\bibinfo {year} {2015})\ pp.\ \bibinfo {pages}
  {110--123},\ \bibinfo {note}
  {\href{http://arxiv.org/abs/1505.03424}{arXiv:1505.03424}}\BibitemShut
  {NoStop}%
\bibitem [{\citenamefont {Bennett}\ \emph {et~al.}(1997)\citenamefont
  {Bennett}, \citenamefont {Bernstein}, \citenamefont {Brassard},\ and\
  \citenamefont {Vazirani}}]{bennett_strengths_1997}%
  \BibitemOpen
  \bibfield  {author} {\bibinfo {author} {\bibfnamefont {C.}~\bibnamefont
  {Bennett}}, \bibinfo {author} {\bibfnamefont {E.}~\bibnamefont {Bernstein}},
  \bibinfo {author} {\bibfnamefont {G.}~\bibnamefont {Brassard}}, \ and\
  \bibinfo {author} {\bibfnamefont {U.}~\bibnamefont {Vazirani}},\ }\bibfield
  {title} {\enquote {\bibinfo {title} {Strengths and {Weaknesses} of {Quantum}
  {Computing}},}\ }\href {\doibase 10.1137/S0097539796300933} {\bibfield
  {journal} {\bibinfo  {journal} {SIAM Journal on Computing}\ }\textbf
  {\bibinfo {volume} {26}},\ \bibinfo {pages} {1510--1523} (\bibinfo {year}
  {1997})}\BibitemShut {NoStop}%
\bibitem [{\citenamefont {Farhi}\ and\ \citenamefont
  {Gutmann}(1998)}]{farhi_analog_1998}%
  \BibitemOpen
  \bibfield  {author} {\bibinfo {author} {\bibfnamefont {Edward}\ \bibnamefont
  {Farhi}}\ and\ \bibinfo {author} {\bibfnamefont {Sam}\ \bibnamefont
  {Gutmann}},\ }\bibfield  {title} {\enquote {\bibinfo {title} {Analog analogue
  of a digital quantum computation},}\ }\href {\doibase
  10.1103/PhysRevA.57.2403} {\bibfield  {journal} {\bibinfo  {journal} {Phys.
  Rev. A}\ }\textbf {\bibinfo {volume} {57}},\ \bibinfo {pages} {2403--2406}
  (\bibinfo {year} {1998})}\BibitemShut {NoStop}%
\bibitem [{\citenamefont {Zalka}(1999)}]{zalka_grovers_1999}%
  \BibitemOpen
  \bibfield  {author} {\bibinfo {author} {\bibfnamefont {Christof}\
  \bibnamefont {Zalka}},\ }\bibfield  {title} {\enquote {\bibinfo {title}
  {Grover's quantum searching algorithm is optimal},}\ }\href {\doibase
  10.1103/PhysRevA.60.2746} {\bibfield  {journal} {\bibinfo  {journal}
  {Physical Review A}\ }\textbf {\bibinfo {volume} {60}},\ \bibinfo {pages}
  {2746--2751} (\bibinfo {year} {1999})}\BibitemShut {NoStop}%
\bibitem [{\citenamefont {Diao}\ \emph {et~al.}(2002)\citenamefont {Diao},
  \citenamefont {Zubairy},\ and\ \citenamefont {Chen}}]{diao_quantum_2002}%
  \BibitemOpen
  \bibfield  {author} {\bibinfo {author} {\bibfnamefont {Zijian}\ \bibnamefont
  {Diao}}, \bibinfo {author} {\bibfnamefont {M.~Suhail}\ \bibnamefont
  {Zubairy}}, \ and\ \bibinfo {author} {\bibfnamefont {Goong}\ \bibnamefont
  {Chen}},\ }\bibfield  {title} {\enquote {\bibinfo {title} {A {Quantum}
  {Circuit} {Design} for {Grover}'s {Algorithm}},}\ }\href {\doibase
  10.1515/zna-2002-0810} {\bibfield  {journal} {\bibinfo  {journal}
  {Zeitschrift f\"{u}r Naturforschung}\ }\textbf {\bibinfo {volume} {57a}},\
  \bibinfo {pages} {701--708} (\bibinfo {year} {2002})}\BibitemShut {NoStop}%
\bibitem [{\citenamefont {Farhi}\ \emph {et~al.}(2000)\citenamefont {Farhi},
  \citenamefont {Goldstone}, \citenamefont {Gutmann},\ and\ \citenamefont
  {Sipser}}]{farhi_quantum_2000}%
  \BibitemOpen
  \bibfield  {author} {\bibinfo {author} {\bibfnamefont {Edward}\ \bibnamefont
  {Farhi}}, \bibinfo {author} {\bibfnamefont {Jeffrey}\ \bibnamefont
  {Goldstone}}, \bibinfo {author} {\bibfnamefont {Sam}\ \bibnamefont
  {Gutmann}}, \ and\ \bibinfo {author} {\bibfnamefont {Michael}\ \bibnamefont
  {Sipser}},\ }\bibfield  {title} {\enquote {\bibinfo {title} {Quantum
  {Computation} by {Adiabatic} {Evolution}},}\ }\href
  {http://arxiv.org/abs/quant-ph/0001106} {\bibfield  {journal} {\bibinfo
  {journal} {arXiv:0001106}\ } (\bibinfo {year} {2000})}\BibitemShut {NoStop}%
\bibitem [{\citenamefont {Isakov}\ \emph {et~al.}(2016)\citenamefont {Isakov},
  \citenamefont {Mazzola}, \citenamefont {Smelyanskiy}, \citenamefont {Jiang},
  \citenamefont {Boixo}, \citenamefont {Neven},\ and\ \citenamefont
  {Troyer}}]{isakov_understanding_2016}%
  \BibitemOpen
  \bibfield  {author} {\bibinfo {author} {\bibfnamefont {Sergei~V.}\
  \bibnamefont {Isakov}}, \bibinfo {author} {\bibfnamefont {Guglielmo}\
  \bibnamefont {Mazzola}}, \bibinfo {author} {\bibfnamefont {Vadim~N.}\
  \bibnamefont {Smelyanskiy}}, \bibinfo {author} {\bibfnamefont {Zhang}\
  \bibnamefont {Jiang}}, \bibinfo {author} {\bibfnamefont {Sergio}\
  \bibnamefont {Boixo}}, \bibinfo {author} {\bibfnamefont {Hartmut}\
  \bibnamefont {Neven}}, \ and\ \bibinfo {author} {\bibfnamefont {Matthias}\
  \bibnamefont {Troyer}},\ }\bibfield  {title} {\enquote {\bibinfo {title}
  {Understanding {Quantum} {Tunneling} through {Quantum} {Monte} {Carlo}
  {Simulations}},}\ }\href {\doibase 10.1103/PhysRevLett.117.180402} {\bibfield
   {journal} {\bibinfo  {journal} {Physical Review Letters}\ }\textbf {\bibinfo
  {volume} {117}},\ \bibinfo {pages} {180402} (\bibinfo {year}
  {2016})}\BibitemShut {NoStop}%
\bibitem [{\citenamefont {Roland}\ and\ \citenamefont
  {Cerf}(2002)}]{roland_quantum_2002}%
  \BibitemOpen
  \bibfield  {author} {\bibinfo {author} {\bibfnamefont {J\'{e}r\'{e}mie}\
  \bibnamefont {Roland}}\ and\ \bibinfo {author} {\bibfnamefont {Nicolas~J.}\
  \bibnamefont {Cerf}},\ }\bibfield  {title} {\enquote {\bibinfo {title}
  {Quantum search by local adiabatic evolution},}\ }\href {\doibase
  10.1103/PhysRevA.65.042308} {\bibfield  {journal} {\bibinfo  {journal}
  {Physical Review A}\ }\textbf {\bibinfo {volume} {65}},\ \bibinfo {pages}
  {042308} (\bibinfo {year} {2002})}\BibitemShut {NoStop}%
\bibitem [{\citenamefont {Radcliffe}(1971)}]{radcliffe_properties_1971}%
  \BibitemOpen
  \bibfield  {author} {\bibinfo {author} {\bibfnamefont {J.~M.}\ \bibnamefont
  {Radcliffe}},\ }\bibfield  {title} {\enquote {\bibinfo {title} {Some
  properties of coherent spin states},}\ }\href {\doibase
  10.1088/0305-4470/4/3/009} {\bibfield  {journal} {\bibinfo  {journal}
  {Journal of Physics A: General Physics}\ }\textbf {\bibinfo {volume} {4}},\
  \bibinfo {pages} {313} (\bibinfo {year} {1971})}\BibitemShut {NoStop}%
\bibitem [{\citenamefont {Perelomov}(1986)}]{perelomov_generalized_1986}%
  \BibitemOpen
  \bibfield  {author} {\bibinfo {author} {\bibfnamefont {Askold}\ \bibnamefont
  {Perelomov}},\ }\href {http://link.springer.com/10.1007/978-3-642-61629-7}
  {\emph {\bibinfo {title} {Generalized {Coherent} {States} and {Their}
  {Applications}}}}\ (\bibinfo  {publisher} {Springer Berlin Heidelberg},\
  \bibinfo {address} {Berlin, Heidelberg},\ \bibinfo {year} {1986})\BibitemShut
  {NoStop}%
\bibitem [{\citenamefont {Young}\ \emph {et~al.}(2013)\citenamefont {Young},
  \citenamefont {Sarovar},\ and\ \citenamefont
  {Blume-Kohout}}]{young_error_2013}%
  \BibitemOpen
  \bibfield  {author} {\bibinfo {author} {\bibfnamefont {Kevin~C.}\
  \bibnamefont {Young}}, \bibinfo {author} {\bibfnamefont {Mohan}\ \bibnamefont
  {Sarovar}}, \ and\ \bibinfo {author} {\bibfnamefont {Robin}\ \bibnamefont
  {Blume-Kohout}},\ }\bibfield  {title} {\enquote {\bibinfo {title} {Error
  suppression and error correction in adiabatic quantum computation: Techniques
  and challenges},}\ }\href {\doibase 10.1103/PhysRevX.3.041013} {\bibfield
  {journal} {\bibinfo  {journal} {Physical Review X}\ }\textbf {\bibinfo
  {volume} {3}},\ \bibinfo {pages} {041013} (\bibinfo {year}
  {2013})}\BibitemShut {NoStop}%
\end{thebibliography}

%

\end{document}